\documentclass[12pt]{article}
\usepackage[dvipsnames]{xcolor}
\usepackage{color}
\usepackage[english]{babel}
\usepackage{amsmath}
\usepackage{amsfonts}
\usepackage{amssymb}
\usepackage{graphicx}
\def\ai{\'{\i}}
\def\lp{\left(}
\def\rp{\right)}

\def\be{\begin{equation}}
\def\ee{\end{equation}}

\begin{document}

\title{\vspace{-1.5cm} \bf Tides and traversability in gravastars and other related geometries}

\author{
C. Tomasini\footnote{e-mail: ctomasini@df.uba.ar} , 
C. Simeone\footnote{e-mail: csimeone@df.uba.ar} , 
and 
E. Rub\'in de Celis\footnote{e-mail: erdec@df.uba.ar} 
\\
{\footnotesize Universidad de Buenos Aires,  Facultad de Ciencias Exactas y Naturales, Departamento de F\ai sica,}\\ 
{\footnotesize 
 Buenos Aires, Argentina and}\\
{\footnotesize 
CONICET - Universidad de Buenos Aires, Instituto de F\ai sica de Buenos Aires (IFIBA),}\\ 
{\footnotesize 
Buenos Aires, Argentina
}}
\date{\small \today}

\maketitle
\vspace{0.6cm} 
\begin{abstract}

Tidal effects related to the traversability across thin shells are examined in spherically symmetric geometries. We focus mainly on shells separating inner from outer regions of gravastars (de Sitter  --i.e. $\Lambda>0$-- interior and Schwarzschild exterior of mass parameter $M$), but we also examine other related geometries by including the possibility of a negative cosmological constant and, besides, non trivial topologies where the shell separates two outer regions.  The analysis is developed for radially traversing objects and for tides in both radial and transverse directions, which present difficulties of somewhat different nature. Transverse tides across shells which satisfy the flare--out condition are the most troublesome, while shells in trivial topologies, i.e. geometries with one asymptotic region, are more indulgent with the issue of large tides. Besides, contradicting other cases analyzed in previous works, we find that large radial tides cannot be avoided when traveling across the shell in the gravastar solution, but in non-trivial topologies they can. We study with special attention the traversability in practice of the transition layer in the thin-shell gravastar solution. In particular, a finite object which traverses radially the shell in a gravastar with $\sqrt{\Lambda}\ll 1/M$ undergoes a compression effect in both the transverse and the radial directions due to the tides associated to the thin layer. The results are interpreted in terms of the total momentum transfer obtained by integrating the travel time of the object.

\end{abstract}

\section{Introduction}

Tidal forces associated with the traversability of surfaces and transition regions within environments of astrophysical or theoretical relevance have been the subject of study since the first interpretations of the Schwarzschild metric and its event horizon. An observer travels safely across the surface of the event horizon if the mass parameter $M$ is large enough given that tides scale with $M^{-2}$. Traversability issues have been widely addressed in black holes and quasi-black hole backgrounds as well as other compact objects or wormholes. Particularly, the effects of tides play an important role in the characterization of wormhole geometries \cite{Morris:1988cz,book} and, with the current renewed interest in such non-trivial solutions, are especially considered in the theoretical construction of traversable throats  (\cite{Maldacena:2020sxe,Emparan:2020ldj,Rueda:2022wge,Avalos:2022inm,Estrada:2023pny,Kavya:2023tjf}, references therein and thereof). Nevertheless, many times, authors refer to a theoretical traversability across throats regarding only the absence of horizons, but forget to study traversability in practice given the possibility of withstanding tidal forces. In this sense, we have recently addressed the analysis of traversability in practice for thin-shell type geometries, either in wormholes with two asymptotic regions \cite{nos21,nos22}, as well as through shells in asymptotically trivial geometries \cite{nos23}, and now -in this work- we study tidal effects in trivial and non-trivial topologies, but with special attention to the traversability in practice of the transition layer in the thin-shell gravastar solution.

The gravastar solution to Einstein's field equations was suggested more than twenty years ago as an alternative model to the final state of stellar collapse that could prevent the appearance of event horizons \cite{mazmot}. By eliminating the singularity and information paradox problems present in black holes, it received much attention due to its conceptual simplicity and advantages when taking quantum effects into account. Gravastars are essentially spherically symmetric spacetimes with an inner de Sitter region and an outer Schwarzschild geometry joined by a matter transition layer. Therefore, these solutions can be classified according to the type of layer at the juncture: one model uses continuous matter with an anisotropic pressure \cite{cat,DeBenedictis:2005vp}, and the other is constructed with infinitely thin shells in combination or not with a finite thickness transition region \cite{Mazur:2004fk,MartinMoruno:2011rm,viswil}. In any case, they are considered compact objects that could be an alternative to black holes in the sense that they share the same exterior geometry up to the transition layer placed just before the appearance of the event horizons.

Even before Mazur and Mottola coined the name gravastar in \cite{mazmot} and \cite{Mazur:2004fk}, Chapline et al. \cite{chaphoh} had developed a similar prototype whose fundamental idea is that a phase transition analogous to the one that occurs at the critical point of the Bose--Einstein condensate occurs at the horizon of a black hole. In the original works \cite{mazmot,Mazur:2004fk}, the authors proposed that gravastars can be the endpoint of gravitational collapse as a thermodynamically stable solution, with a globally defined Killing time, and consistent with quantum theory. Soon after the original works, Visser and Wiltshire constructed a simplified version of the Mazur and Mottola model applying the thin-shell formalism of Israel and Lanczos \cite{isr,lanc}, using only one shell, and studied the mechanical stability against spherically symmetric perturbations to find some equations of state for the matter layer that lead to stable gravastars \cite{viswil}.
On the other hand, Cattoen et al. \cite{cat} developed a continuous thick-layer model showing that gravastars cannot be perfect fluids and hence anisotropic pressure should be included. Being an alternative to black holes, gravastars are an interesting and intriguing background which have been studied to try to understand their formation process and characterize the type of matter present at the transition layer, as well as to study any physical process from what its geometry implies. The thin-shell model of a gravastar vas studied dinamically by Rocha et al. \cite{Rocha:2008yd}, and many other models have been analyzed: changing the interior and exterior geometries \cite{Carter:2005pi}, testing stability against totally inelastic collisions with a dust shell and generalized equation of state \cite{Gaspar:2010bs}, considering a radiating shell \cite{Chan:2011wi}, including generalized dark energy interiors \cite{lobo}, and including electromagnetic field \cite{Horvat:2008ch}. The thin-shell model has been the preferred background to test gravastar applications and implications as astrophysical objects in the last years; for example, gravitational lensing effects were computed using gravastars to compare the images they produce with the ones obtained from black holes geometries as candidates for the massive compact objects observed in nature \cite{Sakai:2014pga,Kubo:2016ada}. The observation of gravitational waves by LIGO has broken down a barrier to accessing some evidence of this model that could be tested with the improved sensitivity of Advanced LIGO, Virgo and other detectors such as LISA \cite{Pani:2009hk}. 

As mentioned before, the shell characterizes the matter layer which separates the inner de Sitter and outer Schwarzschild geometries but it has never been studied regarding the tidal effects it produces on test objects traversing it.  In what follows we are going to study this issue for the thin-shell gravastar solution.  Given the relation between curvature and tides expressed in terms of the Riemann tensor, whenever a model implying curvature jumps is applied, we should pay attention to the consequences on objects traversing near or across the associated peculiar matter distributions. The approach of infinitely thin layers in the description of gravastars involves jumps of the extrinsic curvature tensor; hence we consider that the study of tidal forces constitutes an important topic to be taken into account in order to achieve a more complete knowledge of these hypothetical objects. We will perform a detailed analysis of tides in both radial and transverse directions, which present difficulties of somewhat different nature. We focus mainly on shells separating inner from outer regions of gravastars (de Sitter  --i.e. $\Lambda>0$-- interior and Schwarzschild exterior), but we also examine other related geometries by including the possibility of a negative cosmological constant and, besides, non trivial topologies where the shell separates two outer regions. The article is organized as follows: In section 2 we review the general treatment and formulation of the proper relative acceleration undergone by a test object as it traverses the shell. In section 3 we use Lanczos equations to relate the extrinsic curvature jumps with the matter at the shell and their general contribution to tidal accelerations for the generic static spherically symmetric metric. In section 4 we present the gravastar geometry and the related geometries we will work with: gravastar and wormholes are analyzed separately in 4.1 and 4.2, respectively, and the summarized results are presented in 4.3. Final considerations with general conclusions are discussed in section 5, where the tides are interpreted by expressing them in terms of relative velocities of the object traveling through the layer.

\section{General treatment}

The starting point of our analysis are static geometries of  the form
\be\label{metric1}
ds_{\pm}^2 = g^{\pm}_{00}\, dt_{\pm}^2 + g^{\pm}_{rr}\, dr_{\pm}^2 + g^{\pm}_{\theta\theta}\, d\theta^2 + g^{\pm}_{\varphi\varphi}\, d\varphi^2 \,
\ee
at each side of an infinitely thin shell of constant radius $r_\pm = b_\pm$. 
In such spacetimes we study the tidal forces on an object passing through the shell.  Tides  are proportional to the relative covariant acceleration between two points of the object: 
\be\label{A}
(\Delta a)^\mu =
- g^{\rho\mu}{R}_{\rho\alpha\nu\beta} V^\alpha (\Delta x)^\nu V^\beta,
\ee   
where $V^\mu$ are the components of its four-velocity, $(\Delta x)^\nu$ is the oriented separation between the two points, and ${R}_{\rho\alpha\nu\beta}$ is the Riemann tensor. Let us define the normal coordinate associated to the radial direction: for an inner region connected to an outer one it  is given by  $d\eta = + \sqrt{g_{rr}^{\pm}}
dr_{\pm}$, and the
shell is located at $\eta = \eta_0$; for the case of two asymptotic regions we have
 $d\eta = \pm\sqrt{g_{rr}^{\pm}}
dr_{\pm}$, which measures the perpendicular proper distance near the throat placed 
at $\eta= 0$. The unit normal vector to the
shell is then $\eta_\mu = \partial_\mu\eta$, pointing from the side $(-)$ to the other side $(+)$. Therefore the Riemann tensor can be written as \cite{book,grav}
\begin{eqnarray} \label{riemann}
R_{\mu\alpha\nu\beta} 
 & = & \Theta(-\eta+\eta^*) R^{-}_{\mu\alpha\nu\beta} + \Theta(\eta-\eta^*) R^{+}_{\mu\alpha\nu\beta} \nonumber\\& & 
- \, \delta(\eta-\eta*) 
\left[\kappa_{\alpha\beta} \, n_\mu n_\nu+\kappa_{\mu\nu} \, n_\alpha n_\beta-\kappa_{\alpha\nu} \, n_\mu n_\beta-\kappa_{\mu\beta} \, n_\alpha n_\nu\right] 
\end{eqnarray}
with $\eta^*=0$ for two asymptotic regions and $\eta^*=\eta_0$ for trivial topologies. $R^{\mp}_{\mu\alpha\nu\beta}$ is the regular part of the tensor at each side,  and
\be
\kappa_{\alpha\beta} = \frac{1}{2} \left(\frac{\partial g^+_{\alpha\beta}}{\partial \eta}\Big|_{b_+} - \frac{\partial g^-_{\alpha\beta}}{\partial\eta}\Big|_{b_-} \right)
\ee
is the jump of the extrinsic curvature across the shell. In what follows we will restrict to spherical symmetry and objects moving radially across the shell.  The proper magnitude of the relative acceleration for a radially extended object across the shell, is \cite{nos21}
\be\label{A2}
\Delta a_r=-
{\kappa^{0}}_{0}+\left[  {{R^{r0}}_{r0}}^{-}\Big|_{b_-} +{{R^{r0}}_{r0}}^{+}\Big|_{b_+} \right]
\frac{\Delta \tilde\eta}{2} 
+ \mathcal{O}(\Delta \tilde\eta^2) 
,
\ee
where $\Delta\tilde\eta$ is the proper radial separation between the two points; if the object considered is at rest, then  $\Delta\tilde\eta=\Delta \eta$. The contribution associated with the curvature jump component ${\kappa^{0}}_0$ is independent of the separation because an integration over the extension $\Delta \tilde{\eta}$ of the object has been done to be able to interprete the Dirac delta function coming from the last term in the Riemann tensor (\ref{riemann}). We emphasize that the previous expression is valid to study relative accelerations in objects that extend radially from one side to the other of the shell, that is, assuming objects larger than the hypothetical thickness of the shell.
Objects smaller than shell thickness cannot be analyzed within the thin shell model; to study objects smaller than the actual thickness of a shell, we need a thick shell model.

The regular part of the Riemann tensor in each smooth region in the vicinity of the shell can be put as 
\begin{equation}
{R^{r0}}_{r0}  =  -\,\frac{1}{4g_{rr}g_{00}}\left\{ 2g_{00,rr}-g_{00,r}\frac{\lp g_{00}g_{rr}\rp_{,r}}{g_{00}g_{rr}}\right\}
\end{equation}
and, in general, it does not pose a serious problem in what regards traversability (see for instance Ref. \cite{martin} and also the related work \cite{genc}). 

For transverse objects, oriented in an ortogonal direction to the radial trajectory, the proper relative acceleration in the transverse direction is \cite{nos21}:
\begin{eqnarray}  \label{Ap}
\Delta a_\perp & = & \Delta x_\perp\, \frac{\gamma \beta \,
{\kappa^{\varphi}}_{\varphi}}{\delta\tau}\nonumber\\
&  & +\,
\frac{\Delta x_{\perp}}{2} \,
\left[
{{R^{\varphi 0}}_{\varphi 0}}^-
+\gamma^2 \beta^2 \,
 \left(
 {{R^{\varphi 0 }}_{\varphi 0}}^- -  \, {{R^{\varphi r}}_{ \varphi r}}^-
\right)
\right]_{b_-}\nonumber\\
& & +\,\frac{\Delta x_{\perp}}{2}\left[{{R^{\varphi 0}}_{\varphi 0}}^+ +\gamma^2 \beta^2 \,
 \left(
 {{R^{\varphi 0 }}_{\varphi 0}}^+ -  \, {{R^{\varphi r}}_{ \varphi r}}^+
\right)
\right]_{b_+},
\end{eqnarray}
with $\Delta x_\perp$ the oriented transverse proper separation between the two points and $\delta\tau$ the (infinitely short) proper time in which the transverse object passes through the infinitely thin shell.
As usual, $\beta$ is the radial speed of the object as measured in an orthonormal frame at rest at the vicinities of the throat and  $\gamma=1/\sqrt{1-\beta^2}$. 
To obtain the latter expression, the relative acceleration has been averaged over the interval $\delta \tau$ to interprete the Dirac delta function appearing in (\ref{riemann}).
The first term in (\ref{Ap}) is ideally divergent unless the component ${\kappa^{\varphi}}_{\varphi}$ could be made to vanish \cite{nos22}, or if the object remains at rest extended along the shell. The regular contributions of the Riemann tensor are
\begin{eqnarray}
{R^{\varphi 0}}_{\varphi 0} & = & -\,\frac{g_{\varphi\varphi,r}g_{00,r}}{4g_{rr}g_{00}g_{\varphi\varphi}},\\
{R^{\varphi r}}_{\varphi r}& = & {R^{\varphi 0}}_{\varphi 0} -\frac{1}{4g_{rr}g_{\varphi\varphi}}\left\{ 2g_{\varphi\varphi,rr}-g_{\varphi\varphi,r}\frac{\lp g_{00}g_{\varphi\varphi}g_{rr}\rp_{,r}}{g_{00}g_{\varphi\varphi}g_{rr}}\right\}.
\end{eqnarray}
Of course, completely analogous expressions hold for the angle $\theta$ and the components  ${R^{\theta 0}}_{\theta 0}$ and ${R^{\theta r}}_{\theta r}$ of the curvature tensor.

\section{Tides and shell matter}

The mathematical description of ideally infinitely thin layers of matter is given by the  Lanczos equations \cite{lanc,isr,sen,darm}, which relate the energy surface density and pressure on a shell and the jump of the first derivatives of the metric; they read
\be  \label{le2}
8\pi {S^i}_j  = \kappa\, {\delta^i}_j - {\kappa^i}_j
\ee 
where ${S^i}_j$  is the surface stress  tensor and $\kappa$ is the trace of ${\kappa^i}_j$,  the jump in the extrinsic curvature tensor. In the case of spherical symmetry treated here ${S^i}_j=(-\sigma,p,p)$ and ${\kappa^\varphi}_\varphi={\kappa^\theta}_\theta$; then
\begin{eqnarray}
\sigma & = & -\frac{1}{4\pi}{\kappa^\varphi}_\varphi,\label{en}\\
p & = & \frac{1}{8\pi}\lp{\kappa^0}_0+{\kappa^\varphi}_\varphi\rp.\label{pres}
\end{eqnarray}
According to Eqs. (\ref{A2}) and (\ref{Ap}) the relation between the shell's contributions to the tidal accelerations --let us denote them as ${\cal T}_r$ and ${\cal T}_\perp$-- and the jump of the extrinsic curvature is given by
\begin{eqnarray}
{\cal T}_r 
& = &
- {\kappa^0}_0,\\
{\cal T}_\perp 
& = &
\Delta x_{\perp} \lp \frac{\gamma \beta}{\delta \tau} \rp  {\kappa^\varphi}_\varphi  .
\end{eqnarray}
We find that the contribution implying possible difficulties in the transverse directions cannot vanish unless the shell energy density  $\sigma$ is null; thus it would happen only for matter with peculiar properties. In fact, a vanishing energy density together with just one negative pressure would mean the violation of the null, weak and strong energy conditions, that is, a shell constituted by {\it exotic} matter. Instead, the issue with the contribution of the shell to the radial tide can be straightforwardly avoided if ${\kappa^0}_0=0$, which does not imply by itself any undesirable restriction on the character of the shell matter. Start from  spherically symmetric geometries 
\be
ds_\pm^2=- f_\pm(r_\pm)dt_\pm ^2+ g_\pm(r_\pm)dr_\pm^2+ h_\pm(r_\pm)(d\theta^2+\sin^2\theta d\varphi^2),\label{metric}
\ee
connected by a spherical shell placed at the surface $\Sigma$ defined by $r_\pm=b_\pm$. The relation between $b_-$ and $b_+$ comes from the condition of continuity of the geometry across $\Sigma$, which with the choice of equal angular coordinates at both sides gives  $h_-(b_-)=h_+(b_+)\equiv h_\Sigma$. The continuity also implies $f_-(b_-)dt_-^2=f_+(b_+)dt_+^2$, and on the surface $\Sigma$ we have the induced $(2+1)$-dimensional metric 
$ds_\Sigma^2=-dt_\Sigma^2+h_\Sigma(t_\Sigma)(d\theta^2+\sin^2\theta d\varphi^2)$
 where $t_\Sigma$ is the proper time on the shell. The curvature jump is given by 
\be
{\kappa^0}_0 =  \frac{f_+'(b_+)}{2f_+(b_+)\sqrt{g_+(b_+)}}-\delta\frac{f_-'(b_-)}{2f_-(b_-)\sqrt{g_-(b_-)}},\label{koo}
\ee
\be 
{\kappa^\theta}_\theta =   {\kappa^\varphi}_\varphi =\frac{h_+'(b_+)}{2h_+(b_+)\sqrt{g_+(b_+)}}-\delta\frac{h_-'(b_-)}{2h_-(b_-)\sqrt{g_-(b_-)}}.\label{kpp} 
\ee
where a prime denotes a derivative with respect to $r_{\pm}$; we introduce $\delta$ to distinguish the case of only one asymptotic region  $(\delta=+1)$ from the case of two asymptotic regions  $(\delta =-1)$ --a wormhole if the flare-out condition $h'_{\pm}(b_\pm)>0$ is satisfied. This  comes from the different signs relating the normal coordinate $\eta$ with the radial coordinates $r_\pm$ for each kind of topology: from the definitions of he previous section it is clear that for trivial topologies it is $\mathrm{sgn}(d\eta)=\mathrm{sgn}(dr_\pm)$, while for geometries with two asymptotic regions we have  $\mathrm{sgn}(dr_+)=\mathrm{sgn}(d\eta)$ and $\mathrm{sgn}(dr_-)=-\mathrm{sgn}(d\eta)$. We should note that, in a strict sense, the sign  $\delta=-1$  could also correspond to only one exterior region connected on the shell to an inner region like the so-called  {\it bag of gold} or also {\it baby universe} \cite{wheeler}; this would require a  non monotonic $h_-(r_-)$ function. However, we are not interested in such configurations, so they are not included in our analysis.
  
\section{Gravastars and wormholes} 

The whole spacetime associated to a gravastar can be described, in a first approximation, by an outer  Schwarzschild geometry joined, on a spherical shell of ideally null thickness, to a de Sitter inner submanifold \cite{viswil}:
\be
ds_\pm^2=-f_\pm(r)dt_\pm^2+f_\pm^{-1}(r)dr^2+r^2\left(d\theta^2+\sin^2 \theta d\varphi^2\right),
\ee
\be
f_+(r)=1-\frac{2M}{r},
\ee
\be
f_-(r)=1-\frac{\Lambda r^2}{3}.
\ee
Here $M>0$ is the usual ADM mass and $\Lambda > 0$; the choice $\Lambda<0$ would account for the anti-de Sitter geometry, which we will also briefly consider. The continuity of the geometry across the shell at the joining surface  determines $b_-=b_+=b$. This surface is placed just outside the Schwarzschild radius $2M$ \cite{viswil}, then we can take $b/(2M)=\mu \gtrsim 1$; besides, to avoid the de Sitter cosmological horizon, $\Lambda$ and $M$ must allow the choice  $b< r_{C}=\sqrt{3/\Lambda}$. If we also consider the case of a shell connecting  the outer Schwarzschild geometry with {\it another exterior} region of the de Sitter form, then, instead of a gravastar, we would be dealing with a wormhole geometry and in one side we would have to manage, in one way or another, the problem of the cosmological horizon (see below).

\subsection{Gravastars}

For only one asymptotic region, that is an inner de Sitter submanifold joined to an outer Schwarzschild region, equations (\ref{koo}) and (\ref{kpp}) with $\delta=+1$ give the components of the jump of the extrinsic curvature 
\be
{\kappa^0}_0= \frac{M}{b^2\sqrt{1-2M/b}}+\frac{\Lambda b}{3\sqrt{1-\Lambda b^2/3}},
\ee 
\be
{\kappa^\varphi}_\varphi= \frac{1}{b}\lp\sqrt{1-2M/b}-\sqrt{1-\Lambda b^2/3}\rp.
\ee 
It is easy to verify that for the gravastar  the jump  ${\kappa^\varphi}_\varphi$ associated with divergent transverse tides vanishes with the choice $6M/\Lambda= b^3$, which in the approximation adopted corresponds to
\be\label{cfifi}
\Lambda M^2=\frac{3}{4\mu^3}\lesssim \frac{3}{4}.
\ee
The condition ${\kappa^0}_0=0$ which avoids from the start any radial tides problem, instead, cannot be fulfilled for positive values of the cosmological constant. We must remark that the vanishing troublesome contribution to the transverse tide could be achieved only at the price of three physically somewhat undesirable features: First, ${\kappa^\varphi}_\varphi=0$ directly implies a shell with vanishing energy density. Second,  the relation $b=2M\mu$ together with (\ref{cfifi}) leads to
\be
{\kappa^0}_0=\frac{3}{4M\mu^{3/2}\sqrt{\mu-1} }
\ee
which under the assumption $\mu \gtrsim 1$ implies a very large radial tide across the shell; moreover, this tide diverges in the Mazur--Mottola limit $\mu\to 1$ in which all the mass of the gravastar is attributed to the
energy of the de Sitter ``vacuum''(see Refs. \cite{viswil} and \cite{mazmot}). Besides, according to Eq. (\ref{pres}),  this has the consequence of a very large pressure (which diverges in the same limit) on the shell. And third, for such relation between the cosmological constant and the mass, we would have $r_C=2\mu^{3/2}M$; though this is larger than the transition radius $b=2\mu M$ and then there would be no cosmological horizon in the complete spacetime, the condition $\mu\gtrsim 1$ determines a configuration with a shell placed very near the would-be troublesome horizon of the inner geometry.

If the parameter range is extended to admit  configurations with $\Lambda<0$ (anti-de Sitter interior geometry \cite{viswil,Carter:2005pi}, which does not present the problem of a cosmological horizon), then the jump component ${\kappa^\varphi}_\varphi$ cannot vanish, while the component ${\kappa^0}_0$ would vanish with the choice
 \be\label{c002}
\Lambda M^2=-\frac{3}{2\mu^2\lp -1+\sqrt{1+16\mu(\mu-1)}\rp}.
\ee
Associated with such relation between the parameters, we would have a curvature jump corresponding to the transverse directions given by
\be
{\kappa^\varphi}_\varphi=\frac{1}{2M\mu}\lp\sqrt{\frac{\mu-1}{\mu}}-\sqrt{1+\frac{2}{-1+\sqrt{1+16\mu(\mu-1)}}}\rp.
\ee
For $\mu\gtrsim 1$ we have ${\kappa^\varphi}_\varphi<0$, and then the energy density would be positive and the pressure would be negative.

\subsection{Wormholes}

We can also write down the expressions of the extrinsic curvature jump for the case of a shell connecting  the outer Schwarzschild geometry with {\it another exterior} region of the de Sitter form; then, instead of a gravastar, we would have a wormhole and from equations (\ref{koo}) and (\ref{kpp}) with $\delta=-1$ we obtain
\be\label{kw0}
{\kappa^0}_0= \frac{M}{b^2\sqrt{1-2M/b}}-\frac{\Lambda b}{3\sqrt{1-\Lambda b^2/3}},
\ee 
\be\label{kwp}
{\kappa^\varphi}_\varphi= \frac{1}{b}\lp\sqrt{1-2M/b}+\sqrt{1-\Lambda b^2/3}\rp.
\ee
It is clear from (\ref{kw0}) and (\ref{kwp}) that, comparing with a gravastar, for a wormhole the situation is reversed: the component  ${\kappa^\varphi}_\varphi$ of the extrinsic curvature cannot vanish, but it is possible to achieve the vanishing condition on the $00$ component avoiding crossing radial tides: we obtain ${\kappa^0}_0=0$ for 
\be\label{c00}
\Lambda M^2=\frac{3}{2\mu^2\lp 1+\sqrt{1+16\mu(\mu-1)}\rp}.
\ee
As above, we express the result in an analogous way in terms of the adimensional product $\Lambda M^2$; a second root is also obtained  from ${\kappa^0}_0=0$ but it lacks physical meaning\footnote{Note that the product associated to improving radial tides for a  wormhole is slightly smaller than the one resulting from the condition on the transverse tides for a gravastar. Both conditions converge to $\Lambda M^2=3/4$ for the limit $\mu\to 1$.}. Under the restriction given by Eq. (\ref{c00}), for the transverse directions we obtain 
\be
{\kappa^\varphi}_\varphi=\frac{1}{2M\mu}\lp\sqrt{\frac{\mu-1}{\mu}}+\sqrt{1-\frac{2}{1+\sqrt{1+16\mu(\mu-1)}}}\rp
\ee
and correspondingly
\be
\sigma  =  -\frac{1}{4\pi}{\kappa^\varphi}_\varphi,\ \ \ \ \  
p  = -\frac{\sigma}{2},
\ee
so the energy density is negative (as expected) and the pressure is positive. Another aspect to be considered in the wormhole case is the location of the cosmological horizon at $r_C=\sqrt{3/\Lambda}$: with a little algebra for conveniently rearranging the expressions, the relation (\ref{c00}) leads to
\be
r_C=2\mu M\sqrt{\frac{1}{2}+\sqrt{\frac{1}{4}+4\mu(\mu-1)}}.
\ee
This is quite unsatisfactory, as it means $r_C\gtrsim 2 M \mu=b$, that is a cosmological horizon very near the wormhole throat (understanding proximity in terms of the scale given by the corresponding radius $b$); this could be avoided, of course, with the introduction of a second shell between $b=2M\mu $ and $r_C=\sqrt{3/\Lambda}$ allowing for an asymptotically flat or Schwarzschild geometry, but it is easy to verify that under the assumption $\mu\gtrsim 1$ this would add a large jump of the component $00$ of the extrinsic curvature, with the consequent problem in the radial tide across the new shell. On the other hand, in what regards wormhole geometries,  the anti-de Sitter case is of very limited interest, as it makes impossible to cancel any of the  components of the extrinsic curvature jump. 

\subsection{Smooth nearby regions and tides}

In all cases, it is of interest the relation between the conditions avoiding troublesome contributions to  tides exactly across the shells with the consequent implications regarding the curvature and tides in the immediate smooth regions at both sides ($r_\pm\to b$). We must therefore evaluate the Riemann tensor components ${{R^{r0}}_{r0}}^\pm $ and  $ {{R^{\varphi 0}}_{\varphi 0}}^\pm=
{{R^{\varphi r}}_{\varphi r}}^\pm$.
By replacing the functions  $f_\pm$ we obtain:
\be
{{R^{r0}}_{r0}}^+  =\frac{2M}{b^3},\ \ \ \ \ {{R^{r0}}_{r0}}^-  =\frac{\Lambda}{3},
\ee
\be
{{R^{\varphi 0}}_{\varphi 0}}^+=
{{R^{\varphi r}}_{\varphi r}}^+=-\frac{M}{b^3},\ \ \ \ \ {{R^{\varphi 0}}_{\varphi 0}}^-=
{{R^{\varphi r}}_{\varphi r}}^-=\frac{\Lambda}{3}.
\ee
For the Schwarzschild region, with the adopted model of $b/(2M)=\mu\gtrsim 1$, we have
\be
{{R^{r0}}_{r0}}^+  =\frac{1}{4M^2\mu^3},\ \ \ \ \ 
{{R^{\varphi 0}}_{\varphi 0}}^+=
{{R^{\varphi r}}_{\varphi r}}^+=-\frac{1}{8M^2\mu^3}.
\ee
in any case (gravastar or wormhole, any sign of $\Lambda$). For gravastars ($\Lambda >0$), under the only admissible condition on  the extrinsic curvature jump ${\kappa^\varphi}_\varphi=0$, we would have
\be
{{R^{r0}}_{r0}}^-  =\frac{1}{4M^2\mu^3},\ \ \ \ \ {{R^{\varphi 0}}_{\varphi 0}}^-=
{{R^{\varphi r}}_{\varphi r}}^-=\frac{1}{4M^2\mu^3},
\ee
while with the same global topology but $\Lambda < 0$, the admissible condition on the extrinsic curvature jump is ${\kappa^0}_0=0$ and then
\be
{{R^{r0}}_{r0}}^-  =
{{R^{\varphi 0}}_{\varphi 0}}^-=
{{R^{\varphi r}}_{\varphi r}}^-=-\frac{1}{2M^2\mu^2\lp -1+\sqrt{1+16\mu(\mu-1)}\rp}.
\ee
For wormholes, in the 
case $\Lambda >0$ allowing for  
the vanishing component  
${\kappa^0}_0=0$, we obtain
\be
{{R^{r0}}_{r0}}^-  =
{{R^{\varphi 0}}_{\varphi 0}}^-=
{{R^{\varphi r}}_{\varphi r}}^-=\frac{1}{2M^2\mu^2\lp 1+\sqrt{1+16\mu(\mu-1)}\rp}.
\ee
Consequently, for gravastars with ${\kappa^\varphi}_{\varphi}=0$, tides are proportional to the relative proper accelerations
\begin{eqnarray}
\Delta a_r & = & -\frac{3}{4M\mu^{3/2} \sqrt{\mu-1}}+\frac{\Delta\tilde\eta}{4M^2\mu^3},\\
\Delta a_\perp & = & \frac{\Delta x_\perp}{16M^2\mu^3}.
\end{eqnarray}
The radial tide  is very large for $\mu\gtrsim 1$  and diverges in the limit $\mu\to 1$.  
For the same topology but with anti-de Sitter interior, and with ${\kappa^0}_0=0$, we have
\begin{eqnarray}
\Delta a_r 
& = &
\frac{\Delta\tilde\eta}{8M^2\mu^3}\lp 1-\frac{2\mu}{-1+\sqrt{1+16\mu(\mu-1)}}\rp.
\end{eqnarray}
The radial tide vanishes for infinitely close points, is finite for $\mu>1$ and diverges in the limit $\mu \to 1$. The situation is analogous for a wormhole with one de Sitter side: the difficulties in the transverse tides cannot be straightforwardly avoided, but for ${\kappa^0}_0=0$ we have 
\begin{eqnarray}
\Delta a_r
& = & 
\frac{\Delta\tilde\eta}{8M^2\mu^3}\lp 1+\frac{2\mu}{1+\sqrt{1+16\mu(\mu-1)}}\rp,
\end{eqnarray}
with $\Delta a_r$ vanishing for $\Delta\tilde\eta\to 0$, and equal to $\Delta\tilde\eta/(4M^2\mu^3)$ in the limit $\mu\to 1$.

\section{Discussion}
The behaviour of transverse tides reflects the conclusions in \cite{nos23}, where geometries with only one asymptotic region were considered more indulgent with the issue of large tides:  for a  de Sitter ($\Lambda > 0$) submanifold joined to a Schwarzschild one, the divergence can be cured
in the gravastar configuration (at the price, though, of somewhat ill behaved matter on the shell and extremely large radial tides), while it cannot be avoided for the wormhole geometry; the latter is because 
at both sides of the shell, in configurations with two asymptotic regions, the areas of spheres increase as the corresponding radial coordinate grows and hence both $h'_-$ and $h'_+$  in Eq. (\ref{kpp}) have the same sign. Instead, the situation with radial tides, that is that troublesome contributions can be straightforwardly avoided for the case of two asymptotic regions and not for the trivial topology, may seem to contradict the main line of the discussion in Ref. \cite{nos23}. However, it is a consequence of the behaviour of the de Sitter geometry, with a component $g_{00}$ which {\it decreases} for {\it increasing} values of the radial coordinate, thus yielding a negative value of $f_-'(b)$ together with a positive $f_+'(b)$ in Eq. (\ref{koo}). In fact, to clarify this point a qualitative analysis could be carried out in terms of the preliminary considerations of Ref. \cite{nos21}. Start from the acceleration of a point particle at each side of the shell separating the inner and the outer region of a gravastar:  the radial acceleration of a rest particle
just outside is $a^r=-M/b^2$ (towards the center), while the radial acceleration of a rest particle just inside is $a^r = \Lambda b/3$ (pointing outwards); therefore the relative acceleration cannot vanish and we should not be surprised by the result obtained here. 

A possible compromise solution can be proposed if we relax the approach of infinitely thin shell and consider a more realistic interpretation of the results in terms of layers of finite, though very little, thickness. Then the curvature jumps are to be understood as an estimate of a sort of mean Riemann tensor for a region of  non vanishing thickness corresponding to the location of the matter layer; the associated relative accelerations should be interpreted consequently. Let us focus in the physically most appealing configuration which is the gravastar with inner de Sitter region, and consider a very little cosmological constant; then in terms of a scale given by the Schwarzschild mass, we now assume the relation $\sqrt{\Lambda}\ll 1/M $ which, reasonably, pushes very far away the location of the would-be cosmological horizon before the cut and paste mathematical construction. Then for the curvature jump we obtain 
\begin{eqnarray}
{\kappa^0}_0 & \gtrsim & \frac{1}{4M\mu^{3/2}\sqrt{\mu-1}},\\
{\kappa^\varphi}_\varphi 
& \gtrsim & 
- \frac{1}{2M\mu} \lp 1 - \sqrt{1-\mu^{-1}}\rp.
\end{eqnarray}
Thus we would have a jump of the radial component smaller (by a factor 3) than the one obtained above, which would mean an obvious improvement by itself; both components are regulated by the Schwarzschild mass. To proceed further with the analysis, recall the tidal accelerations associated to the shell's contribution
\begin{eqnarray}
{\cal T}_r \label{Tra} 
& \lesssim & -\frac{1}{4M\mu^{3/2}\sqrt{\mu-1}},\\
{\cal T}_{\perp} \label{Tpa}
& \gtrsim & 
- \frac{\gamma\beta}{2M \mu\, \delta\tau} 
\lp 1 - \sqrt{1-\mu^{-1}}\rp
\Delta x_\perp,
\end{eqnarray}
where $\delta\tau$ is the traversing time across the shell of, now, non vanishing though little
 thickness $\delta\epsilon$; this time in the denominator would be non null and then it would not necessarily imply a problem with traversability. The negative signs in both expressions represent a transverse and radial compression, respectively. The situation would now  depend, in general, on the mass $M$ and, for the transverse tide, also on the traversing speed of the object; this can be best understood by recalling that under this approach we can define $\gamma \beta = \delta \epsilon /\delta\tau$,
in terms of the proper time and thickness\footnote{Recall that the definition of speed would be $\beta = \delta \tilde{\epsilon}/\delta\tau$, in terms of the proper time and proper thickness $\delta \tilde{\epsilon}$ as measured by the traversing object, and the Lorentz contraction establishes $\delta{\epsilon} = \gamma  \delta \tilde{\epsilon}$.} , to write
\be
{\cal T}_{\perp}  \gtrsim  - \frac{\gamma^2 \beta^2}{2M\mu \, \delta{\epsilon}}\lp1 - \sqrt{1-\mu^{-1}} \rp \Delta x_\perp	\,,
\ee
which contains the typical $\gamma^2$ dependence of transverse tides (see for example the smooth parts in (\ref{Ap}), or the analysis in \cite{book}). It is clear that the absolute magnitude of the transverse tides are reduced by lowering the speed $\beta$ of the crossing through the shell. 

An alternative interpretation of the above results can be made in terms of the proper relative velocities obtained by multiplying relative accelerations by the corresponding time interval. This is equivalent to consider the total momentum transfer obtained by integrating over the travel time. This approach allows to definitively get rid of the non-smooth expressions and calculate finite physical magnitudes to interprete the tidal effects associated to traversing the shell. Lets, for example, define the transverse relative velocity $\Delta v_{\perp} = \delta \tau\Delta a_{\perp} $ associated to the two extremes of a transverse object, which would represent a finite change of the internal energy if the object is allowed to react as it travels across the shell. Multiplying (\ref{Tpa}) by the traversing time $\delta \tau$ we have
\be
\Delta  v_{\perp} 
\gtrsim
- \frac{\gamma\beta}{2M \mu} 
\lp 1 - \sqrt{1-\mu^{-1}}\rp
\Delta x_\perp \,.
\ee
If $\beta \to 0$ the relative velocity is zero and the static case is recovered, while $\beta \to 1$ reproduces the expected divergence with $\gamma$ for transverse effects \cite{book} (unbearable compression).
Analogously, if we define the time it takes for a radially extended object with proper separation $\Delta \tilde{\eta}$ to travel across the shell as $\delta \tau_r = \Delta \tilde{\eta} / \beta$, we can obtain the radial relative velocity change $\Delta v_r = \delta \tau_r \Delta a_r $ due to the passage through the shell given by 
\be
\Delta v_r \lesssim 
- \frac{\Delta\tilde{\eta}}{4M \mu^{3/2}\beta \sqrt{\mu-1}} \,.
\ee
The latter is now proportional to the proper separation at the price of the dependence on $\beta$; if $\beta \to 0$ the magnitude diverges because the object remains at rest across the shell during an infinite time and, if it is allowed to react, the relative velocity decreases until a final compression, while if $\beta \to 1$ the relative velocity is finite, as it must be, given that the radial effects are finite for finite size of the object and finite traveling time interval, independently of the traveling speed. In conclusion, a finite object which traverses radially the transition layer of a thin-shell gravastar with $\sqrt{\Lambda}\ll 1/M $ suffers a compression effect in both the transverse and the radial directions due to the tides associated to the thin layer.

\section*{Acknowledgement}

Part of this work is framed in and supported by the projects ``Proyecto de Investigaci\'on Plurianuales" PIP 2022-2024 (112202101 00225CO)
and ``Proyecto UBACyT 2023 Modalidad II" (20020220400140BA).


\begin{thebibliography}{99} 

\bibitem{Morris:1988cz} M. S. Morris and K. S. Thorne, Am. J. Phys. \textbf{56}, 395 (1988).

\bibitem{book} M. Visser, \textit{Lorentzian Wormholes} (AIP Press, New York, 1996).

\bibitem{Maldacena:2020sxe} J. Maldacena and A. Milekhin, Phys. Rev. D \textbf{103}, 066007 (2021).


\bibitem{Emparan:2020ldj} R. Emparan, B. Grado-White, D. Marolf and M. Tomasevic,
JHEP \textbf{05}, 032 (2021).

\bibitem{Rueda:2022wge} A. Rueda, R. Avalos and E. Contreras, Eur. Phys. J. C \textbf{82}, 605 (2022).

\bibitem{Avalos:2022inm} R. Avalos and E. Contreras, Ann. Phys. \textbf{446}, 169128 (2022).

\bibitem{Estrada:2023pny} M. Estrada and C. R. Muniz, JCAP \textbf{03}, 055 (2023).

\bibitem{Kavya:2023tjf} N. S. Kavya, V. Venkatesha, G. Mustafa, P. K. Sahoo and S. V. D. Rashmi, Chin. J. Phys. \textbf{84}, 1 (2023).

\bibitem{nos21} E. R. de Celis and C. Simeone, Eur. Phys. J. C \textbf{81}, 937 (2021).

\bibitem{nos22} E. R. de Celis and C. Simeone, Eur. Phys. J. C \textbf{82}, 1035 (2022).

\bibitem{nos23} E. R. de Celis and C. Simeone, Eur. Phys. J. C \textbf{83}, 863 (2023).

\bibitem{mazmot} P. O. Mazur and E. Mottola, Universe \textbf{9}, 88 (2023).

\bibitem{cat} C. Cattoen, T. Faber and M. Visser, Class. Quantum Grav. \textbf{22}, 4189 (2005).

\bibitem{DeBenedictis:2005vp} A. DeBenedictis, D. Horvat, S. Ilijic, S. Kloster and K. S. Viswanathan, Class. Quantum Grav. \textbf{23}, 2303 (2006).

\bibitem{Mazur:2004fk} P. O. Mazur and E. Mottola, ``Gravitational vacuum condensate stars,'' Proc. Nat. Acad. Sci. \textbf{101}, 9545 (2004).

\bibitem{MartinMoruno:2011rm} P. Martin Moruno, N. Montelongo Garcia, F. S. N. Lobo and M. Visser, JCAP \textbf{03}, 034 (2012).

\bibitem{viswil} M. Visser and D. L. Wiltshire, Class. Quantum Grav. \textbf{21}, 1135 (2004).

\bibitem{chaphoh} G. Chapline, E. Hohlfeld, R. B. Laughlin and D. I. Santiago, Int. J. Mod. Phys. A \textbf{18}, 3587 (2003).

\bibitem{isr} W. Israel, Nuovo Cimento B \textbf{44}, 1 (1966); \textbf{48}, 463(E) (1967).

\bibitem{lanc} K. Lanczos, Ann. Phys. (Leipzig) \textbf{379}, 518 (1924).

\bibitem{Rocha:2008yd} P. Rocha, A. Y. Miguelote, R. Chan, M. F. da Silva, N. O. Santos and A. Wang, JCAP \textbf{06}, 025 (2008).

\bibitem{Carter:2005pi} B. M. N. Carter, Class. Quantum Grav. \textbf{22}, 4551 (2005).

\bibitem{Gaspar:2010bs} M. E. Gaspar and I. Racz, Class. Quantum Grav. \textbf{27}, 185004 (2010).

\bibitem{Chan:2011wi} R. Chan, M. F. A. da Silva, J. F. V. da Rocha and A. Wang, JCAP \textbf{10}, 013 (2011).

\bibitem{lobo} F. S. N. Lobo, Class. Quantum Grav. \textbf{23}, 1525 (2006).

\bibitem{Horvat:2008ch} D. Horvat, S. Ilijic and A. Marunovic, Class. Quantum Grav. \textbf{26}, 025003 (2009).

\bibitem{Sakai:2014pga} N. Sakai, H. Saida and T. Tamaki, Phys. Rev. D \textbf{90},  104013 (2014).

\bibitem{Kubo:2016ada} T. Kubo and N. Sakai, Phys. Rev. D \textbf{93}, 084051 (2016).

\bibitem{Pani:2009hk} P. Pani, E. Berti, V. Cardoso, Y. Chen and R. Norte, J. Phys. Conf. Ser. \textbf{222}, 012032 (2010).

\bibitem{grav} S. Weinberg, {\it Gravitation and Cosmology} (John Wiley and sons, New York, 1972).

\bibitem{martin} M. G. Richarte and C. Simeone, Int. J. Mod. Phys. D {\bf  17}, 1179 (2008).

\bibitem{genc} O. Gen\c{c},  Int. J. Mod. Phys. D {\bf 32}, 2350014 (2023).

\bibitem{sen} N. Sen, Ann. Phys. (Leipzig) \textbf{378}, 365 (1924).

\bibitem{darm} G. Darmois, M\'{e}morial des Sciences Math\'{e}matiques, Fascicule XXV, Chap. 5  (Gauthier-Villars, Paris, 1927).

\bibitem{wheeler} J. Wheeler, {\it Geometrodynamics and the Issue of the Final State, in  Relativity, Groups, and Topology, 1963 Les Houches Lectures} (B. S. DeWitt and C. M. DeWitt, eds., Gordon and Breach, New York, 1964).







\end{thebibliography}
\end{document}